\documentclass[%
reprint,
superscriptaddress,
amsmath,
amssymb,
aps,
pra,
]{revtex4-2}
\usepackage[english]{babel}
\usepackage{graphicx}
\usepackage{dcolumn}
\usepackage{bm}
\usepackage[colorlinks=true,
            linkcolor=cyan,
            citecolor=cyan,
            urlcolor=black]{hyperref}
\usepackage[mathlines]{lineno}
\usepackage{orcidlink}
\usepackage{blindtext}
\usepackage{changes}
\usepackage{layouts}
\usepackage{braket}
\usepackage{bbold}
\usepackage{mathrsfs}
\usepackage{xcolor}
\usepackage{mathtools}
\usepackage{placeins}

\usepackage[normalem]{ulem}

\definechangesauthor[name={Author},color=blue]{au}
\setaddedmarkup{\textcolor{green!60!black}{#1}}
\setdeletedmarkup{\textcolor{red!100!black}{\sout{#1}}}
\definecolor{mplC1}{RGB}{255,127,14}
\setcommentmarkup{\textcolor{mplC1}{#1}}

\definechangesauthor[name={Simone}, color=red]{A}
\definechangesauthor[name={Nicola}, color=blue]{N}

\newcommand{\bb}[1]{\boldsymbol{#1}}

\begin{document}

\preprint{APS/123-QED}

\title{
Atomic correlation effects in collapse-induced spontaneous radiation
}

\author{Simone Manti\,\orcidlink{0000-0003-3770-0863}}
\thanks{Mail: \texttt{Simone.Manti@lnf.infn.it} (Corresponding Author)}
\affiliation{Laboratori Nazionali di Frascati INFN, Frascati, Italy}

\author{Nicola Bortolotti\,\orcidlink{0000-0002-9344-3657}}
\affiliation{Centro Ricerche Enrico Fermi, Museo Storico della Fisica e Centro Studi e Ricerche "Enrico Fermi", Roma, Italy}
\affiliation{Laboratori Nazionali di Frascati INFN, Frascati, Italy}

\author{Lajos Diósi\,\orcidlink{0000-0003-4722-3220}}
\affiliation{Wigner Research Center for Physics, H-1525 Budapest 114, P.O.Box 49, Hungary}
\affiliation{Eötvös Loránd University, H-1117 Budapest, Pázmány Péter stny, 1/A, Hungary}

\author{Kristian Piscicchia\,\orcidlink{0000-0001-6879-452X}}
\affiliation{Centro Ricerche Enrico Fermi, Museo Storico della Fisica e Centro Studi e Ricerche "Enrico Fermi", Roma, Italy}
\affiliation{Laboratori Nazionali di Frascati INFN, Frascati, Italy}

\author{Catalina Curceanu\,\orcidlink{0000-0002-1990-0127}}
\affiliation{Laboratori Nazionali di Frascati INFN, Frascati, Italy}
\affiliation{IFIN-HH, Institutul National pentru Fizica si Inginerie Nucleara Horia Hulubei, 30 Reactorului, 077125, Magurele, Romania}


\date{\today}

\begin{abstract}
Collapse models introduce stochastic and nonlinear modifications in the quantum dynamics, predicting observable effects, such as spontaneous radiation from charged particles, which can be used to constrain their parameters. Recently, attention has focused on the 1–100 keV energy range, where the wavelength of the emitted photons becomes comparable to atomic dimensions, making the emission sensitive to atomic structure and leading to model-dependent behaviors that enable their discrimination. Here, we derive a general expression for the spontaneous emission rate for arbitrary noise, providing a framework that systematically incorporates the atomic structure through the radial distribution of the emitters, modulated by the specific collapse model. The formalism recovers previous results in the appropriate limits and naturally includes new low-energy effects, such as cancellation mechanisms arising from charge correlations. We evaluate the rates for germanium and xenon within the Di\'osi–Penrose and Continuous Spontaneous Localization models, showing how these correlations modify the predicted emission rates. This approach provides a unified framework to account for atomic effects and enables more robust, material-dependent experimental constraints on collapse model parameters.
\end{abstract}

\maketitle


\section{Introduction}
The collapse of the wavefunction remains one of the most debated aspects of quantum mechanics \cite{Gao_2018,bassiModelsWavefunctionCollapse2013d}. Standard quantum theory describes measurement-induced reduction as an external postulate, leaving open the question of whether wavefunction collapse is a fundamental physical process or an effective description \cite{schlosshauerDecoherenceMeasurementProblem2005} Spontaneous collapse models (SCMs) address this issue by introducing stochastic and nonlinear modifications of the Schr\"odinger equation, providing a unified dynamical framework in which localization emerges spontaneously and universally \cite{bassiCollapseModelsTheoretical2014}.
Several SCMs have been proposed, differing in their physical motivations and mathematical structure. The Ghirardi--Rimini--Weber model \cite{ghirardiUnifiedDynamicsMicroscopic1986} introduces discrete localization events, while its continuous extension, the Continuous Spontaneous Localization (CSL) model \cite{pearle1999csl,diosiModelsUniversalReduction1989a}, describes a diffusive process driven by a classical noise field. The Di\'osi--Penrose (DP) model \cite{diosiGravitationQuantummechanicalLocalization1984,penroseGravitysRoleQuantum1996}, on the other hand, connects wavefunction collapse to gravitational effects, linking the localization mechanism to the mass distribution of the system. An earlier approach due to K\'arolyh\'azy \cite{karolyhazyGravitationQuantumMechanics1966} attributes wavefunction collapse to intrinsic fluctuations of the space-time metric, providing a gravitationally motivated source of decoherence; this framework has recently attracted renewed attention \cite{bortolottiExperimentalExclusionGeneralized2026,figuratoTestabilityKarolyhazyModel2024}, motivating further theoretical and phenomenological investigations. These models are typically characterized by collapse parameters that set the strength ($\lambda$) and spatial resolution ($\sigma$) of the localization process, defining the scale at which quantum superpositions are effectively suppressed \cite{Feldmann_2012}.
Experimental tests of SCMs have been pursued through a variety of methods, including interferometric experiments \cite{nimmrichterTestingSpontaneousLocalization2011}, bulk heating measurements \cite{adlerBulkHeatingEffects2018}, cantilever-based setups \cite{vinanteUpperBoundsSpontaneous2016}, and searches for excess noise \cite{carlessoMultilayerTestMasses2018}, each probing different aspects of the underlying stochastic dynamics. Among the non-interferometric approaches \cite{carlessoPresentStatusFuture2022}, particular attention has been devoted to the search for electromagnetic emission induced by the collapse noise \cite{fu1997spontaneous}. Indeed, the stochastic field responsible for localization imparts a diffusive motion to charged particles, leading to the emission of radiation \cite{bassi2009electromagnetic}. This spontaneous radiation constitutes one of the most direct experimental signatures, providing a powerful avenue to constrain the SCM parameters \cite{adler2013spontaneous,piscicchiaCSLCollapseModel2017}.
Experimental searches have therefore focused on detecting this evanescent radiation in ultra-low-background environments, where the expected signal can be disentangled from conventional sources. In this context, underground experiments employing high-purity detectors have achieved some of the most stringent bounds. Notable examples include the VIP experiment \cite{donadi2021novel,donadiUndergroundTestGravityrelated2021,bortolottiExperimentalExclusionGeneralized2026}, the MAJORANA Demonstrator \cite{abgrall2014majorana}, and the XENONnT experiment \cite{aprileXENONnTDarkMatter2024a}, all of which analyze x-ray emission spectra in the keV-MeV range. By exploiting the predicted spectral shape of collapse-induced radiation together with the extremely low environmental background, these experiments are able to set competitive and robust limits on the parameters of SCMs.
Recently, increasing attention has been devoted to the low-energy regime of spontaneous radiation \cite{piscicchiaXRayEmissionAtomic2024b}. While at high energies the emission rate typically follows an inverse-energy scaling, $(N_P^2+N_e)/E$ with $N_P$ and $N_e$ the number of protons and electrons in the system respectively, deviations arise as the photon wavelength becomes comparable to the characteristic size of the emitting system. In this regime, interference and cancellation effects within the atom become relevant, leading to a nontrivial modification of the spectral shape \cite{xenoncollaborationChallengingSpontaneousQuantum2026}. Notably, different SCMs, such as CSL and DP, predict qualitatively distinct behaviors of this cancellation effect at low energies, enhancing the discriminating power of experimental searches \cite{piscicchiaXRayEmissionAtomic2024b}.
A proper description of this regime requires an accurate treatment of atomic correlations, as SCMs effectively modulate the interaction between atomic emitters and the radiation field. Phenomenological models become inadequate when the atomic structure dominates, and an electronic-structure-based approach is needed to disentangle genuine collapse-induced effects from standard atomic contributions \cite{bartellEffectsElectronCorrelation1964}. This enables a consistent and model-independent evaluation of the emission rate, providing a robust basis to constrain SCM parameters.
In this article, we derive and evaluate the spontaneous collapse-induced emission rate for an arbitrary noise field. The resulting expression separates the model-dependent modulation of the emitter–radiation coupling from the atomic contribution, which is fully determined by the distribution of inter-particle separations. This leads to an evaluation based on radial distribution functions (RDFs) of the emitters, computed from first-principles electronic structure calculations. Using this approach, we obtain quantitative predictions for germanium (Ge) and xenon (Xe) and evaluate the emission rates within the CSL and DP models.
The paper is structured as follows: in Sec.~\ref{sec:derivation} we present the rate derivation for arbitrary noise; in Sec.~\ref{sec:evaluation} we outline its evaluation using RDFs, together with the implementation details; in Sec.~\ref{sec:results} we present and discuss results for Ge and Xe for CSL and DP models; and Sec.~\ref{sec:conclusions} presents the conclusions.
%
\section{Emission Rate for arbitrary noise}
\label{sec:derivation}
%
We derive here a general expression for the collapse-induced spontaneous emission rate, valid for arbitrary noise correlations. We focus on mass-proportional SCMs, which employ the mass density operator $\hat\mu(\bb x)$ as the collapse operator, ensuring spontaneous suppression of large quantum fluctuations of the mass density. The dynamics is governed by a master Lindblad equation \cite{lindbladGeneratorsQuantumDynamical1976}, where a non-unitary term modifies the standard Hamiltonian dynamics. This new term depends on the mass density of the quantum system and the noise correlation function, and can be interpreted in terms of spacetime fluctuations \cite{bortolottiFundamentalLimitsClock2025}. The master equation can be derived from a stochastic Schr\"odinger equation,
\begin{equation}\label{eq: stochastic Schroedinger equation}
	\frac{d}{dt}\ket{\psi_t} = -\frac{i}{\hbar} \left[ \hat{H} + \int d^3x {\hat\mu}(\bb{x}) \phi(\bb{x},t) \right] \ket{\psi_t} ,
\end{equation}
where $\hat{H}$ is the system's Hamiltonian and $\phi$ is a fluctuating component of the classical Newtonian potential, with zero average and correlation 
\begin{equation}\label{general correlation}
    \mathbb{E}[\phi(\bb{x},t)\phi(\bb{y},t')] = \mathcal{C}(|\bb{x}-\bb{y}|,|t-t'|) .
\end{equation}
Both CSL and DP models fall into this class of SCMs. For these models the correlation \eqref{general correlation} factorizes into a spatial $\mathcal{D}(r)$ and a temporal $\mathcal{G}(\tau)$ term, with the former given by \cite{bortolottiFundamentalLimitsClock2025}
\begin{subequations}
\begin{align}
	&\mathcal{D}_\text{CSL}(r) = \frac{\hbar^2 \lambda}{m_0^2} e^{-r^2/4\sigma^2}, \label{eq:CSL noise} \\
    &\mathcal{D}_\text{DP}(r) = \frac{\hbar G}{r} \text{erf}\left( \frac{r}{2\sigma} \right) , \label{eq:DP noise}
\end{align}
\end{subequations}
where $m_0$ is a reference mass, typically chosen to be the nucleon mass. The parameter $\lambda$ sets the collapse rate for the CSL model, while $\sigma$ is a smearing length which defines the resolution of the localization process.

Following previous analyses \cite{donadi2021novel,donadiUndergroundTestGravityrelated2021}, we adopt a semiclassical approach to derive the general expression for the emission rate. For energies above 1 keV, this approach has been shown \cite{adler2007lower} to yield results equivalent to those obtained from a fully quantum mechanical treatment. In this framework, the spontaneous radiation rate is computed from the random acceleration of the charged particles within the atoms. According to Eq. \eqref{eq: stochastic Schroedinger equation}, particles experience acceleration due to fluctuations of the Newtonian potential
\begin{equation}
	\ddot{\bb{r}}_i(t) = -\boldsymbol{\nabla} \phi(\bb{r}_i,t) ,
\end{equation}
where $\bb{r}_i(t)$ is the position of the $i$-th particle at time $t$. These accelerations are correlated as
\begin{equation}\label{coordinate space acceleration correlation}
	\mathbb{E}\left[\ddot{\bb{r}}_i^*(t) \otimes \ddot{\bb{r}}_j(t') \right] = \boldsymbol{\nabla}_i \otimes \boldsymbol{\nabla}_j \mathcal{C}(|\bb{r}_i-\bb{r}_j|,|t-t'|) ,
\end{equation}
which follows from Eq. \eqref{general correlation} by using linearity of the average and differentiation. For later convenience, we evaluate the correlation \eqref{coordinate space acceleration correlation} in Fourier space. Using the Fourier transform of the Newtonian potential correlation function $\phi(\bb{r},\omega)$ we have
\begin{equation}
		\mathbb{E}\left[ \phi^*(\bb{r}_i,\omega) \phi(\bb{r}_j,\omega') \right] = 2\pi \delta(\omega-\omega')\mathcal{C}(\bb{r}_i-\bb{r}_j,\omega) ,
\end{equation}
and expressing the derivatives with respect to the positions of particles $i$ and $j$ in terms of their separation vector $\bb{r}_{ij}=\bb{r}_i-\bb{r}_j$, we obtain
\begin{equation}\label{acceleration correlations in fourier}
	\mathbb{E}\left[\ddot{\bb{r}}_i^*(\omega) \otimes \ddot{\bb{r}}_j(\omega') \right] = -2\pi\delta(\omega-\omega') \boldsymbol{\nabla} \otimes\boldsymbol{\nabla} \mathcal{C}(\bb{r}_{ij},\omega) .
\end{equation}
We can now compute the radiation emitted by the accelerated particles. The total power radiated by a system of charged particles through a spherical surface of radius $R$ is
\begin{equation}\label{total power}
        P(t) = \int d\Omega R^2 S(R\hat{\boldsymbol{n}},t) ,
\end{equation}
where $S(R\hat{\bb{n}},t)$ is the magnitude of the Poynting vector $\bb{S} = \bb{E} \times \bb{B} / \mu_0$ at position $\bb{R} = R\hat{\bb{n}}$ and time $t$. For a system of particles, the radiation contribution to the Poynting vector is:
\begin{equation}\label{radiation poynting vector}
    \bb{S}_\text{rad}(\bb{R},t) = \epsilon_0c \sum_{ij} \boldsymbol{E}_i\cdot\boldsymbol{E}_j\hat{\bb{\mathcal{R}}}_j ,
\end{equation}
where $\boldsymbol{\mathcal{R}}_i = c(t-t_r) \hat{\bb{\mathcal{R}}}_i$ is the vector from the retarded position $\bb{r}_i(t_r)$ of the $i$-th particle to the field point $\bb{R}$. We shall assume $\mathcal{R} \gg r$, such that each particle can be approximated at the origin and $\boldsymbol{\mathcal{R}}_i\approx \bb{R}$. The total power \eqref{total power} is related to the emission rate through
\begin{equation}
	P = \int_0^\infty d\omega \hbar\omega \frac{d\Gamma}{d\omega}.
\end{equation}
We therefore write the spectral density of total power in terms of the spectral density $S_{EE}({\bb{R}},\omega)$
\begin{equation}\label{power in frequency}
    \frac{dP}{d\omega} = 2\int d\Omega R^2 S_{EE}({\bb{R}},\omega) , \qquad \omega\geq0,
\end{equation}
which is related to the emission rate by
\begin{equation}\label{rate from power}
	\frac{dP}{d\omega} = \hbar\omega \frac{d\Gamma}{d\omega} .
\end{equation}
Inserting Eq. \eqref{radiation poynting vector} into the total power \eqref{total power} and expanding in Fourier modes, one obtains
\begin{equation}\label{spectral density}
    \frac{\epsilon_0 c}{(2\pi)^2} \bb{E}^*(\bb{R},\omega) \cdot \bb{E}(\bb{R},\omega') = \delta(\omega-\omega') S_{EE}({\bb{R}},\omega) ,
\end{equation}
with $S_{EE}({\bb{R}},\omega)=S_{EE}({\bb{R}},-\omega)$. 

In the nonrelativistic regime, the Fourier mode of the radiation field generated by the $i$-th particle is
\begin{equation}\label{E Fourier}
    \bb{E}_i(\bb{R},\omega) = \frac{q_i}{4\pi\epsilon_0 c^2 R} \int_{-\infty}^\infty dt e^{-i\omega t} \left[ (\hat{\bb{n}}\cdot \ddot{\bb{r}}_i(t_r))\hat{\bb{n}} - \ddot{\bb{r}}_i(t_r) \right] ,
\end{equation}
where the accelerations $\ddot{\bb{r}}_i$ are evaluated at retarded time $t_r$. By replacing $t$ with the retarded time $t_r = t - |\bb{R}-\bb{r}_i(t_r)|/c$ and approximating the retarded position of the $i$-th particle with its averaged position $\bar{\bb{r}}_i$, the right hand side of Eq. \eqref{E Fourier} results to be independent of the retarded time and yields
\begin{equation}
\begin{split}
    \bb{E}_i(\bb{R},\omega) = &\frac{q_i}{4\pi\epsilon_0 c^2 R} e^{-i\frac{\omega}{c}|\bb{R}-\bar{\bb{r}}_i|} \\
    &\times\int_{-\infty}^\infty dt e^{-i\omega t} \left[ (\hat{\bb{n}}\cdot \ddot{\bb{r}}_i(t))\hat{\bb{n}} - \ddot{\bb{r}}_i(t) \right] ;
\end{split}
\end{equation}
furthermore, $|\bb{R}-\bar{\bb{r}}_i| \approx R - \hat{\bb{n}}\cdot\bar{\bb{r}}_i$ and $\omega/c = k$ thus
\begin{equation}\label{Fourier electric field}
    \bb{E}_i(\bb{R},\omega) = \frac{q_i\, e^{-ik(R - \hat{\bb{n}}\cdot\bar{\bb{r}}_i)}}{4\pi\epsilon_0 c^2 R}  \left[ (\hat{\bb{n}}\cdot \ddot{\bb{r}}_i(\omega))\hat{\bb{n}} - \ddot{\bb{r}}_i(\omega) \right] .
\end{equation}
Now, for a time average over the stochastic process, the accelerations are to be interpreted as the expectation values $\mathbb{E}[\ddot{\bb{r}}_i]$. So, let us average Eq. \eqref{radiation poynting vector}.  Using expression \eqref{Fourier electric field} for the radiation field generated by the $i$-th particle and the correlation function \eqref{acceleration correlations in fourier} for the accelerations, leads to
\begin{equation}
\begin{split}
    \mathbb{E}\left[ \bb{E}_i^*(\bb{R},\omega) \cdot \bb{E}_j(\bb{R},\omega') \right] = \frac{2\pi \,q_iq_j}{(4\pi\epsilon_0 c^2 R)^2} e^{ik\hat{\bb{n}}\cdot\bar{\bb{r}}_{ij}} \\
    \times \left[ (\hat{\bb{n}}\cdot \boldsymbol{\nabla})^2 - \nabla^2 \right] \mathcal{C}(r_{ij},\omega) \,\delta(\omega-\omega'),
\end{split}
\end{equation}
where $\bar{\bb r}_{ij} = \bar{\bb r}_i - \bar{\bb r}_j$ and we used $k=k'$ due to the delta function. From Eqs. \eqref{spectral density} and \eqref{power in frequency} then follows 
\begin{equation}\label{power density}
\begin{split}
    \frac{dP}{d\omega} = \frac{2}{32\pi^3 \epsilon_0 c^3} \sum_{i,j} q_iq_j \int d\Omega e^{ik\hat{\bb{n}}\cdot\bb{r}_{ij}} \\
    \times\left[ (\hat{\bb{n}}\cdot \boldsymbol{\nabla})^2 - \nabla^2  \right] \mathcal{C}(r_{ij},\omega) ,
\end{split}
\end{equation}
where the integration runs over the direction $\hat{\bb{n}}$. Note that in Eq. \eqref{power density}, as in the remainder of the analysis, we drop the bar to denote the mean distances $\bar{\bb{r}}_{ij}$. The two integrals in Eq. \eqref{power density} can be evaluated as follows
\begin{equation}
    \int d\Omega e^{ik\hat{\bb{n}}\cdot{\bb{r}}_{ij}} \hat{\bb{n}}\otimes\hat{\bb{n}} = -\frac{1}{k^2} \boldsymbol{\nabla}\otimes\boldsymbol{\nabla}  \int d\Omega e^{ik\hat{\bb{n}}\cdot{\bb{r}}_{ij}} ,
\end{equation}
and
\begin{equation}\label{angular integral}
    \int d\Omega e^{ik\hat{\bb{n}}\cdot{\bb{r}}_{ij}} = 4\pi\frac{\sin k {r}_{ij}}{k r_{ij}} .
\end{equation}
Using the relation \eqref{rate from power} between the spectral power density and the rate, we finally obtain the expression 
\begin{equation}\label{exact rate}
	\frac{d\Gamma}{d\omega} = \frac{1}{4\pi^2 \epsilon_0 c^3\hbar\omega} \sum_{i,j} q_iq_j D_{ij}\mathcal{C}(r_{ij},\omega) ,
\end{equation}
where the differential operator $D_{ij}$ is defined as
\begin{equation}
	D_{ij} = -\left[ \frac{1}{k^2} \nabla_\alpha \nabla_\beta \frac{\sin k {r}_{ij}}{k {r}_{ij}} + \delta_{\alpha\beta} \frac{\sin k {r}_{ij}}{k {r}_{ij}}\right] \nabla_\alpha \nabla_\beta ,
\end{equation}
and summation over repeated spatial indices $\alpha$ and $\beta$ is implied. This expression is valid for noise fields characterized by a generic correlation function $\mathcal{C}(r,\omega)$. Eq. \eqref{exact rate} simplifies in isotropic materials, where the distances $\bb{r}_{ij}$ occur with random orientation. Returning to Eq. \eqref{power density}, we now perform the average over the direction of $\bb{r}_{ij}$ before the angular integration. The averaged value of the differential operator in Eq. \eqref{power density} is
\begin{equation}
    \braket{ [ (\hat{\bb{n}}\cdot \boldsymbol{\nabla})^2 - \nabla^2 ] \mathcal{C}({r}_{ij},\omega) }  =  -\frac{2}{3} \nabla^2\mathcal{C}({r}_{ij},\omega).
\end{equation}
Using Eq. \eqref{angular integral} for the remaining angular integration, together with relation \eqref{rate from power}, the expression for the rate becomes
\begin{equation}\label{amorf bulk expression}
    \frac{d\Gamma}{d\omega} = \frac{1}{6\pi^2 \epsilon_0 c^3 \hbar \omega} \sum_{i,j} q_iq_j \frac{\sin k {r}_{ij}}{k {r}_{ij}} \left( -\nabla^2\mathcal{C}({r}_{ij},\omega) \right) .
\end{equation} 
As we already pointed out, for typical colored noises the general correlation function $\mathcal{C}({r}_{ij},\omega)$ factorizes into spatial and temporal/frequency parts: $\mathcal{C}({r},\omega) = \mathcal{D}({r}) \mathcal{G}(\omega)$. Let us define the function
\begin{equation}
	f(r) \coloneqq \frac{1}{6\pi^2 \epsilon_0 c^3} \left( -\nabla^2\mathcal{D}(r) \right) ,
\end{equation}
which modulates the coupling between the emitters in the rate, with its specific form determined by the collapse model employed. The explicit expressions for the CSL and DP models can be obtained from Eqs. \eqref{eq:CSL noise} and \eqref{eq:DP noise}, which give the spatial correlation functions. Evaluating the Laplacian of these expressions yields
\begin{equation}
	f(r_{ij}) = \frac{1}{6\pi^2 \epsilon_0 c^3}
	\begin{cases}
		\dfrac{6\hbar ^{2} \lambda }{4m_{0}^{2} \sigma ^{3}}\biggl( 1-\dfrac{r_{ij}^{2}}{6\sigma ^{2}}\biggr) e^{-r_{ij}^{2} /4\sigma ^{2}}  &\mathbf{\text{(CSL)}} \\[0.4cm]
        \dfrac{\hbar G}{2\sqrt{\pi } \sigma ^{3}} e^{-r_{ij}^{2} /4\sigma ^{2}}  &\mathbf{\text{(DP)}} 
	\end{cases}
    \label{eq:mod_func}
\end{equation}
Finally, the emission rate for a white noise reads  
\begin{equation}
\frac{d\Gamma}{dE}\Big\vert_\text{white} = \frac{1}{E} \sum_{ij} q_i q_j \frac{\sin(kr_{ij})}{kr_{ij}} f(r_{ij}) ,
\label{eq:rate1}
\end{equation}
where $E = \hbar \omega$ is the energy of the emitted photon with frequency $\omega$. Note that the factors $kr_{ij} = Er_{ij}/\hbar c$ are energy dependent. For a colored noise characterized by correlations exponentially decaying in time, the rate is
\begin{equation}\label{colored rate}
	\frac{d\Gamma}{dE}\Big\vert_\text{colored} = \frac{d\Gamma}{dE}\Big\vert_\text{white} \times \frac{E_c^2}{E_c^2 + E^2} ,
\end{equation}
where the cutoff energy is related to the correlation time by $t_c = \hbar/E_c$.
Eqs. \eqref{eq:rate1} and \eqref{colored rate} coincide with the white and colored noise rates analyzed in Ref. \cite{piscicchiaXRayEmissionAtomic2024b} for the CSL and DP cases.
%
\section{Rate evaluation}
\label{sec:evaluation}
%
In this section we disentangle the atomic-structure contribution in the emission rate (\ref{eq:rate1}) from the collapse-model dependence. The former, governed by the spatial distribution of emitters within the atom, is evaluated from first principles, while the latter is fully encoded in the collapse parameters.
The rate in Eq. (\ref{eq:rate1}) can be evaluated by introducing the RDF $P_{ij}(r)$ of emitters $i$ and $j$, defined such that $P_{ij}(r)dr$ is the probability of finding the two emitters at a separation between $r$ and $r+dr$ \cite{hansenTheorySimpleLiquids2013}. With this definition, the discrete sum over emitter pairs in Eq. (\ref{eq:rate1}) can be recast as a sum of radial integrals of the form
\begin{equation}
\frac{d\Gamma}{dE} = \frac{1}{E} \sum_{ij} q_i q_j \int_{0}^{\infty} dr \, P_{ij}(r) \frac{\sin(kr)}{kr} f(r) ,
\label{eq:sumint}
\end{equation}
In the derivation of Sec.~\ref{sec:derivation}, spherical symmetry is used to perform the angular integration. We therefore evaluate the rates for Ge and Xe within an isolated-atom approximation, retaining only intra-atomic electron--electron and electron--nucleus correlations. This amounts to neglecting condensed-matter effects, such as the crystalline structure of Ge and the liquid phase of Xe. These effects are governed by interatomic correlations on length scales of several angstroms and are expected to provide subleading corrections in the 1--100 keV energy range considered here. In this range the photon wavelength spans from the interatomic scale to well below the atomic scale, so the dominant corrections to the high-energy $1/E$ behavior arise from the atomic radial distribution functions. The inclusion of crystal or liquid correlations is left for future extensions of the formalism.
To expand the sum (\ref{eq:sumint}) over emitter pairs, we use capital indices for nuclei and lower-case indices for electrons. Considering nucleus pairs, and approximating their mutual separations as negligible on the atomic scale, the nucleus--nucleus contribution to the sum can be written as
\begin{equation}
\sum_{IJ} P_{IJ} (r)=N_{P}^{2} \delta (r) ,
\end{equation}
peaked around zero distance from the atom and $N_P$ the number of protons in the atom. For the sum over nuclei and electrons:
\begin{equation}
\sum_{iJ} P_{iJ} (r) = 2N_{P} \sum _{i} 4\pi r^{2}\rho _{i}(r) ,
\end{equation}
where the probability of finding an electron at distance $r$ from the nucleus is given by the radial density distribution $4\pi r^{2}\rho_{i}(r)$ for electron $i$. Finally, for the electron–electron term
\begin{equation}
\sum_{ij} P_{ij}(r) = N_{e} \delta(r) + \sum_{i\neq j} P_{ij}(r) ,
\end{equation}
where the sum is separated into contributions from the same electron and from distinct electron pairs. Here we follow the convention typically used in collapse-model studies and treat the emitters as ordered pairs, to facilitate comparison with previous approximations, like the $(N_P^2+N_e)/E$ behaviour \cite{donadi2021novel}. Therefore all the double sums include diagonal self terms.     Restricting instead to unordered pairs would merely change some combinatorial factors; all qualitative conclusions (e.g., cancellation and modulation effects) are unchanged and are discussed in Appendix~\ref{sec:emitters}.
We can now define the total nucleus-electron RDF, $R(r)$, and the total electron-electron RDF, $P(r)$, as
\begin{equation}
R(r) = \sum_{i} 4\pi r^{2}\rho_{i}(r)
\qquad 
P(r) = \sum_{i\neq j} P_{ij}(r) ,
\label{eq:rdfs}
\end{equation}
with the following normalizations for ordered pairs
\begin{equation}
\int_0^{\infty} dr \, R(r) = N_e \qquad \int_0^{\infty} dr \, P(r) = N_e (N_e - 1) ,
\label{eq:norm}
\end{equation}
and we can now rewrite the emission rate sum (\ref{eq:sumint}) as
\begin{equation}
\frac{d\Gamma}{dE} = \frac{1}{E}\bigl[N_{P}^{2} + N_{e} - 2N_{P} R(k) + P(k)\bigr] ,
\label{eq:rate_final}
\end{equation}
where we have defined the Fourier-like or sinc-transform of the RDFs modulated by the specific collapse envelope function as
\begin{align}
R(k) &= \int_0^{\infty} dr \, R(r)\frac{\sin(kr)}{kr}f(r) , \\
P(k) &= \int_0^{\infty} dr \, P(r)\frac{\sin(kr)}{kr}f(r) .
\label{eq:trans_rdf}
\end{align}
%
RDFs were evaluated for an isolated atom and electrons can be grouped by their electronic orbitals (i.e. 1s, 2s, 2p, ...). The nuclear-electron RDF can be cast with contribution from every electron orbital $\alpha$
\begin{equation}
R(r) = \sum_{\alpha} n_\alpha\,4\pi r^{2}\rho_{\alpha}(r) ,
\label{eq:rdf_nuc_sum}
\end{equation}
where $n_\alpha$ is the number of electrons in orbital $\alpha$, and $4\pi r^{2}\rho_{\alpha}(r)$ is the radial electron density at distance $r$, with $\rho_{\alpha}(r)=|\psi_{\alpha}(r)|^2$ obtained from the wavefunction of orbital $\alpha$. To obtain the orbital wavefunctions, we perform calculations with the density functional theory (DFT) in the local density approximation for an isolated atom using the GPAW software \cite{10.1063/5.0182685}.
\begin{figure}[h!]
    \centering
    \includegraphics[width=1.\columnwidth]{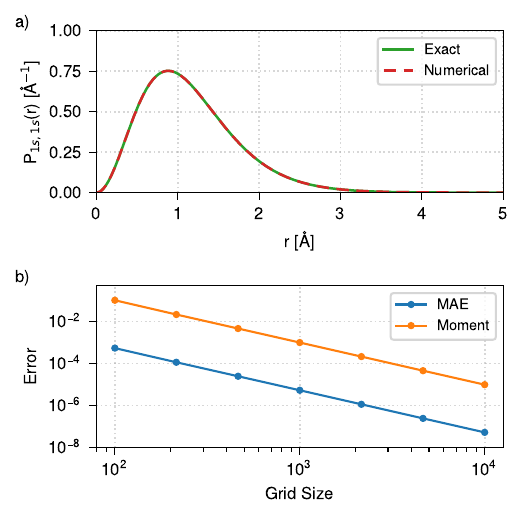}
    \caption{Electron radial distribution function (a) for two 1s electrons using a hydrogenic wavefunction, used to benchmark the implementation against analytical solutions with a grid size of 10\textsuperscript{4} (left). A scan (b) over the grid size is shown on the right, reporting the Mean Absolute Error (MAE) in blue and the first moment in orange, which converges to the analytical value of 35/16 a.u..}
    \label{fig:implementation}
\end{figure}
%
%
The evaluation of the electron--electron RDF is more cumbersome, since correlation effects among electrons naturally introduce additional complications and require further approximations \cite{kingInnerOuterRadial2016}. The sum in (\ref{eq:rdfs}) can be partitioned over pairs of shells $\alpha$ and $\beta$ in a similar fashion
\begin{equation}
P(r) = 2\sum_{\alpha\beta} n_\alpha(n_\beta-\delta_{\alpha\beta})P_{\alpha\beta}(r) ,
\label{eq:rdf_el_sum}
\end{equation}
where $P_{\alpha\beta}(r)$ is the density probability for two electrons in shells $\alpha$ and $\beta$ at distance $r$, also known as intracule integral \cite{coulsonElectronCorrelationGround1961}. It can be evaluated by \cite{beneschDeterminationRadialElectron1972}
\begin{equation}
P_{\alpha\beta}(r) = \int \rho_{\alpha\beta}(\mathbf{r}_1,\mathbf{r}_2)\,\delta(r-r_{12})\,d\mathbf{r}_1\, d\mathbf{r}_2 ,
\label{eq:intracule}
\end{equation}
which requires an approximation for the two-particle density matrix $\rho_{\alpha\beta}(\mathbf{r}_1,\mathbf{r}_2)$. Here, we approximate the two-particle density as a product of single-particle densities, $\rho_{\alpha\beta}(\mathbf{r}_1,\mathbf{r}_2) \approx \rho_{\alpha}(\mathbf{r}_1)\rho_{\beta}(\mathbf{r}_2)$, thereby neglecting exchange and Coulomb correlation effects. The product approximation neglects exchange and Coulomb correlation effects, which mainly modify the short-range part of the electron-pair distribution, but these corrections are not expected to alter the qualitative comparison presented here. Within this approximation, the intracule integral (\ref{eq:intracule}) can be decomposed as \cite{calaisSimpleMethodTreating1962}
\begin{equation}
P_{\alpha\beta}(r) = \sum_{\ell} G_{\alpha\beta}^{(\ell)}I_{\alpha\beta}^{(\ell)}(r) ,
\end{equation}
divided in an angular part $G_{\alpha\beta}^{(\ell)}$ and in a radial part $I_{\alpha\beta}^{(\ell)}$. The angular part is defined in terms of the orbital quantum numbers of the two electrons in terms of the 3j-symbols
\begin{equation}
G_{\alpha\beta}^{(\ell)} = \frac{L_{\alpha} L_{\beta} L}{4\pi }\begin{pmatrix}
l_{\alpha} & l_{\alpha} & l\\
0 & 0 & 0
\end{pmatrix}^{2}\begin{pmatrix}
l_{\beta} & l_{\beta} & l\\
0 & 0 & 0
\end{pmatrix}^{2} ,
\end{equation}
where $L_\alpha = \sqrt{l_\alpha / 4\pi}$. The radial part is defined in terms of the radial single particle wavefunction $u_\alpha(r_1)$ and $u_\beta(r_2)$ of the electrons in the radial integral
\begin{equation}
I_{\alpha\beta}^{(\ell)}(r) = 2\pi r\int_{0}^{\infty} dr_{1} r_1 u_{\alpha}^{2}(r_{1})\int_{r_1-|r|}^{r_1+|r|}dr_2 r_2 u_{\beta}^{2}(r_{2})\Theta_\ell(\cos\omega)
\label{eq:rad_int} ,
\end{equation}
where $\Theta_\ell(\cos\omega)$ are Legendre polynomials accounting for the angular correlations for non-spherical shells ($\ell\neq 0$), with the angle $\omega$ defined by the two-electron geometry
\begin{equation}
\cos\omega = \frac{r_1^2 + r_2^2 - r^2}{2r_1r_2} .
\end{equation}
\begin{figure}[h!]
    \centering
    \includegraphics[width=\columnwidth]{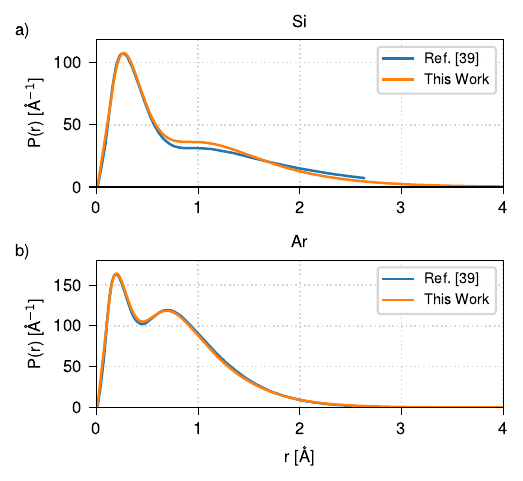}
    \caption{Comparison between the implemented electron-electron RDFs (orange) and reference \cite{QuantumChemistryANU} (blue) for silicon (a) and argon (b).}
    \label{fig:benchmark}
\end{figure}
%
The radial integrals were implemented on a logarithmic grid, with maximum value of the grid as 50 a.u. and the number of grid points chosen to ensure convergence. To benchmark the implementation, we compared the numerical results with analytical cases, which exist only for hydrogenic wavefunctions and for a limited set of ground and excited configurations in the helium atom \cite{boydCoulombHoleExcited1973}. In Fig. \hyperref[fig:implementation]{\ref*{fig:implementation}(a)}, we benchmark our implementation against the analytical result for the 1s hydrogenic wavefunction, for which the analytical radial distribution function is given by \cite{coulsonElectronCorrelationGround1961}
\begin{equation}
P_{1s,1s}(r) =  \frac{r^2}{6}(3 + 6r + 4r^2)e^{-2r} .
\end{equation}
The first moment can be obtained analytically and is equal to 35/16 a.u.. In Fig. \hyperref[fig:implementation]{\ref*{fig:implementation}(b)}, we show the benchmark as a function of the grid size, reporting the mean absolute error (MAE) with respect to the analytical solution and the first moment. A grid size of $10^4$ yields convergence at the $10^{-6}$ level and is therefore used throughout the remainder of this work.
Explicit RDF calculations for Ge and Xe are not available. Many calculations up to Ar are available online at Ref.~\cite{QuantumChemistryANU}. We benchmark the implementation against Si and Ar, for which reference intracule data are available and which share analogous valence-shell structures with Ge and Xe, respectively. In Figs.~\hyperref[fig:benchmark]{\ref*{fig:benchmark}(a)-\ref*{fig:benchmark}(b)}, we compare the resulting electron RDFs with reference data from the literature for silicon, Fig.~\hyperref[fig:benchmark]{\ref*{fig:benchmark}(a)}, and argon, Fig.~\hyperref[fig:benchmark]{\ref*{fig:benchmark}(b)}. Despite the different levels of approximation, using DFT in the present work and ROHF/6-311G Hartree-Fock in Ref.~\cite{gillTwoelectronDistributionFunctions2003}, good agreement is observed, with the first moments agree at ~1\% level, with consistent contributions from the different electronic shells.
\section{Results and Discussion}
\label{sec:results}
\begin{figure}[h!]
    \centering
    \includegraphics[width=1.\linewidth]{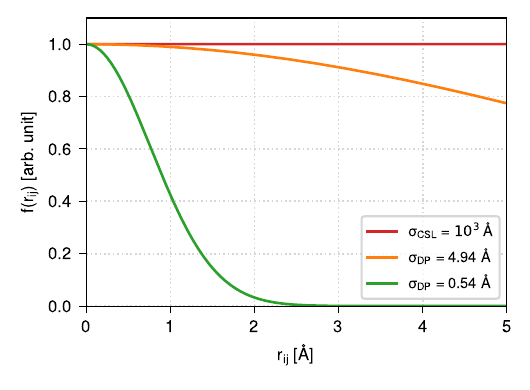}
    \caption{Modulation function $f(r_{ij})$ as a function of the emitters distance $r_{ij}$ for DP model, with values $\sigma_{\mathrm{DP}} = 0.54\,\mathrm{\AA}$ \cite{donadiUndergroundTestGravityrelated2021} (green) and $\sigma_{\mathrm{DP}} = 4.94\,\mathrm{\AA}$ \cite{majoranacollaborationSearchSpontaneousRadiation2022} (orange), and CSL with $\sigma_{\mathrm{CSL}} = 10^3\,\mathrm{\AA}$ \cite{piscicchiaCSLCollapseModel2017} (red).}
    \label{fig:mod_func}
\end{figure}

\begin{figure}[h!]
    \centering
    \includegraphics[width=\columnwidth]{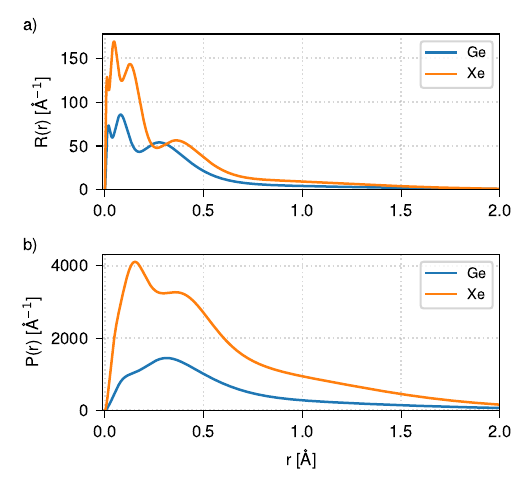}
    \caption{Nuclear-electron $R(r)$ (a) and electron-electron $P(r)$ (b) RDFs for Ge (blue) and Xe (orange) as functions of the interparticle distance $r$.}
    \label{fig:rdf_ge_xe}
\end{figure}

\begin{figure*}[t]
    \centering
    \includegraphics[width=\linewidth]{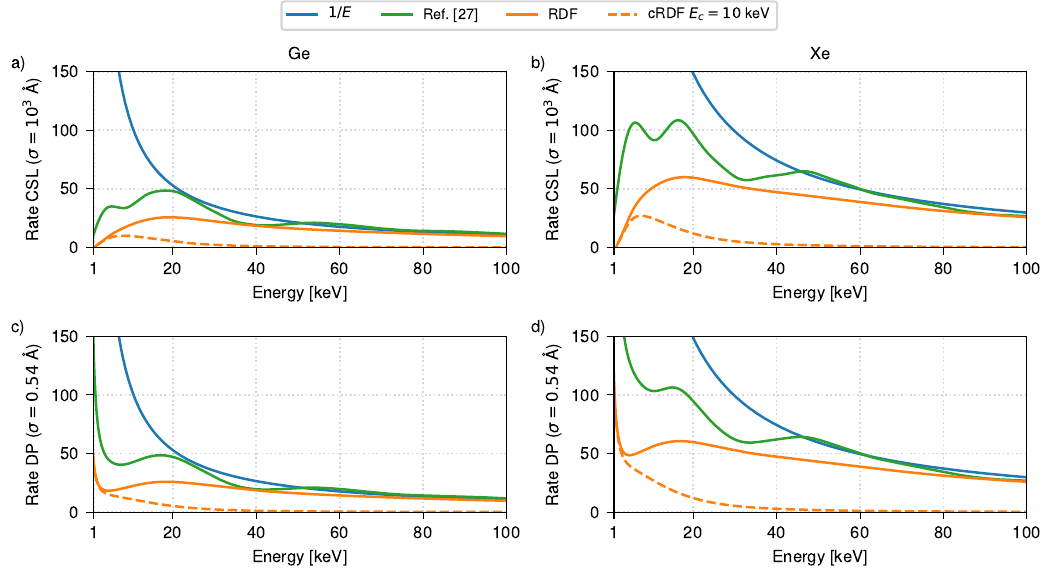}
    \caption{Emission rates for Ge (a) and (c) and Xe (b) and (d) in the 1-100 keV range. The (a) and (b) panels show the RDF rate for CSL with $\sigma$ = $10^3$ \AA \, and the (c) and (d) panels for DP $\sigma$ = 0.54 \AA. For each panel the RDF rate is compared to the 1/E behaviour, the approximation in Ref. \cite{piscicchiaXRayEmissionAtomic2024b} and colored-noise (cRDF) with a cutoff energy of $E_c=10\;\mathrm{keV}$. All rates are shown in arbitrary units.}
    \label{fig:rates_ge_xe}
\end{figure*}

%
Based on the methodology developed in the previous sections, we evaluate here the spontaneous emission rates for Ge and Xe for the CSL and DP models.
The first aspect that distinguishes the behavior of CSL and DP is the modulation function of the emitter, $f(r_{ij})$. For radiation wavelengths on typical atomic distances ($\sim$ \AA), the modulation induced by the two models differs significantly for typical values of the correlation lengths. To show this aspect in Fig. ~\ref{fig:mod_func} we plot the $f(r_{ij})$ for DP and CSL for typical values of their correlation lengths without prefactors in Eq. (\ref{eq:mod_func}). In the DP model, the $f(r_{ij})$ strongly depends on the correlation length over the atomic scale, as illustrated by the change from the VIP value $\sigma_{\mathrm{DP}}=0.54 \,\mathrm{\AA}$ \cite{donadi2021novel} to the value given by MAJORANA $\sigma_{\mathrm{DP}} = 4.94\,\mathrm{\AA}$ \cite{majoranacollaborationSearchSpontaneousRadiation2022}. By contrast, for the CSL model the standard correlation length, $\sigma_{\mathrm{CSL}}=10^3\,\mathrm{\AA}$ \cite{piscicchiaCSLCollapseModel2017}, is several orders of magnitude larger than atomic distances, leaving the modulation function essentially unchanged.
%
This different behavior has profound consequences on the spontaneous collapse emission rate. Since the two models probe atomic length scales differently, this leads to distinct low-energy limits. In the CSL case, a cancellation effect emerges as $E \to 0$, as first discussed in Ref.~\cite{piscicchiaXRayEmissionAtomic2024b}. Within the RDF formalism, this effect arises naturally and corresponds to the absence of spatial modulation between emitter pairs, namely $f(r)=1$. In this limit, using the normalization properties of the transforms defined in Eqs.~(\ref{eq:trans_rdf}) and (\ref{eq:norm}), the rate formula in Eq.~(\ref{eq:rate_final}) reduces to:
\begin{equation}
\lim_{E \to 0}\frac{d\Gamma}{dE} = \frac{1}{E}\bigl[N_{P}^{2} + N_{e} - 2N_{P}N_e + N_e(N_e-1)\bigr] ,
\end{equation}
and for a neutral atom, $N_P = N_e = Z$:
\begin{equation}
\lim_{E \to 0}\frac{d\Gamma}{dE} = \frac{1}{E}\bigl[Z^{2} + Z - 2Z^{2} + Z(Z-1)\bigr] = 0 ,
\end{equation}
which shows that the rate vanishes as the energy approaches zero. The same cancellation effect is recovered using the alternative prescription for counting particle pairs, as discussed in Appendix ~\ref{sec:emitters}. 
This aspect was first introduced in Ref.~\cite{piscicchiaXRayEmissionAtomic2024b}, which captured the qualitative difference between the CSL and DP models. In that work, the sum over emitters was evaluated using a simplified representation of the atomic structure, where electrons were treated as clamped at the mean radius of their respective orbitals.
%
%
Here, we overcome this approximation by explicitly incorporating realistic radial densities of the emitters through the RDFs. The resulting RDFs for Ge and Xe are shown in Figs. \hyperref[fig:rdf_ge_xe]{\ref*{fig:rdf_ge_xe}(a)-\ref*{fig:rdf_ge_xe}(b)} . In both cases, the distributions extend over the angstrom scale. The observed peaks reflect the inclusion of additional orbitals in the sums of Eqs.~(\ref{eq:rdf_nuc_sum}) and (\ref{eq:rdf_el_sum}). As expected, the electronic RDFs appear more diffuse, owing to the broader spatial distribution of the electron cloud.
%
For Ge and Xe, the RDFs were evaluated using nonrelativistic wavefunctions in the radial integrals of Eqs.~(\ref{eq:rad_int}). Owing to the relatively large atomic numbers of these elements, especially Xe, we assessed the sensitivity of the RDFs to relativistic effects by comparing with calculations based on relativistic wavefunctions. The resulting differences remain at the percent level, indicating that relativistic corrections provide a minor contribution within the present framework. We therefore retain the nonrelativistic wavefunctions throughout this work.
Having obtained the RDFs for Ge and Xe, we evaluate the emission rate in Eq.~(\ref{eq:rate_final}) for the DP and CSL models. In Figs.~\hyperref[fig:rates_ge_xe]{\ref*{fig:rates_ge_xe}(a)-\ref*{fig:rates_ge_xe}(d)}, the resulting emission rates are compared with the standard $1/E$ scaling and the clamped-electron approximation of Ref.~\cite{piscicchiaXRayEmissionAtomic2024b} for both Ge and Xe. We also include the colored-noise prediction (cRDF), obtained from Eq.~(\ref{colored rate}) with a cutoff energy of $E_c=10\;\mathrm{keV}$. This value is chosen to allow a direct comparison with the colored-noise benchmark considered in Ref. \cite{piscicchiaXRayEmissionAtomic2024b}.
At high energies, the RDF approach naturally recovers the $1/E$ scaling. This can be understood from the fact that, in this regime, correlations among emitters become ineffective in modulating the radiation. The RDF contributions oscillate rapidly, so their net effect averages out in Eq.~(\ref{eq:rate_final}), leaving the atom unable to significantly modify the emission spectrum.
As the energy decreases, correlations between pairs of emitters begin to play a non-negligible role, leading to a modulation of the rate. Because the RDF approach incorporates realistic radial densities, it predicts a rate that remains consistently below the $1/E$ limit. By contrast, the clamped-electron approximation develops pronounced oscillations, in some regions even exceeding the $1/E$ behavior. These oscillations can be traced back to the artificial localization of the radial density at average orbital distances, which introduces spurious characteristic frequencies in the Fourier transform of the RDFs. In the RDF approach, the smoother and more extended radial densities distribute the spectral weight over a broader range of frequencies, washing out such artificial oscillatory features and yielding a more physical emission rate.
%
While the approximation introduced in Ref. \cite{piscicchiaXRayEmissionAtomic2024b} captures the qualitative low-energy behavior of the CSL and DP models, the RDF-based treatment provides a more realistic description of the spatial distribution of emitters within the atom. This leads to a reduced emission intensity in the low-energy limit. Moreover, the RDF-based approach naturally remains below the asymptotic 1/E behavior, whereas modeling the emitters as clumped point charges introduces unphysical oscillations in the predicted emission rate.
%
%
%
\section{Conclusions}
\label{sec:conclusions}
%
In this work we have derived a general expression for collapse-induced spontaneous radiation valid for arbitrary noise correlations, and reformulated it in a compact pair-sum structure where the roles of collapse dynamics and atomic structure are disentangled. The model dependence is entirely encoded in a spatial modulation kernel, while the atomic contribution enters through the distribution of inter-particle separations, naturally described in terms of RDFs. This formulation provides a systematic and first-principles framework to include atomic structure effects in the evaluation of emission rates. By computing RDFs from electronic wavefunctions, we improve upon previous models in which emitters are approximated as point-like or clamped at their mean orbital radii, providing a more realistic description of their spatial distribution within the atom. As shown in the results section, this leads to quantitatively different predictions, especially in the low-energy regime where the radiation wavelength becomes comparable to atomic scales. A key outcome is the proper treatment of cancellation effects at low energies. For neutral systems, the formalism automatically reproduces the vanishing of the rate in the limit $E \to 0$, while also capturing deviations from the simple $1/E$ scaling at finite energies. The RDF-based approach suppresses unphysical oscillations present in simplified models and yields smoother, physically grounded spectra. We have applied the method to Ge and Xe, relevant targets for low-background experiments, and compared predictions for the CSL and DP models. The analysis highlights how different collapse mechanisms lead to distinct modulations of the emission rate at atomic length scales, enhancing the potential of low-energy measurements in the 1--100~keV range to discriminate between models. The present framework provides the theoretical basis for future analyses of low-background x-ray experiments, enabling material-dependent predictions and more robust inference of collapse model parameters from experimental data.
%
\section{Code Availability Statement}
%
The code used to evaluate the radial distribution functions and spontaneous emission rates presented in this work is publicly available at \url{https://github.com/simonemanti/collapse}.
%
\section{Data Availability Statement}
%
The data used to reproduce the figures for the radial distribution functions and spontaneous emission rates presented in this work are available on Zenodo at \url{https://doi.org/10.5281/zenodo.21234978}, embargo periods may apply.
\begin{acknowledgments}
The authors gratefully acknowledge Riccardo Gargana for providing access to computational resources used in this work. This publication was made possible through the support of Grant No. 62099 from the John Templeton Foundation. The opinions expressed in this publication are those of the authors and do not necessarily reflect the views of the John Templeton Foundation. We acknowledge support from the Foundational Questions Institute and Fetzer Franklin Fund, a donor advised fund of Silicon Valley Community Foundation (Grants No. FQXi-RFP-CPW-2008 and No. FQXi-MGB-2011), and from the INFN (VIP). L.D. was supported by the National Research, Development and Innovation Office “Frontline” Research Excellence Program (Grant No. KKP133827). C.C. and L.D. benefited from the EU COST Actions CA23115 and CA23130. N.B. and K.P. acknowledge support from the Centro Ricerche Enrico Fermi—Museo Storico della Fisica e Centro Studi e Ricerche “Enrico Fermi” (Open Problems in Quantum Mechanics project).
\end{acknowledgments}
%
\vskip10pt
\appendix
\section{Counting emitters in the emission rate}
\label{sec:emitters}
Here we derive the rate formula using the alternative prescription for counting particle pairs. In this approach, the nuclear, nucleus-electron, and electron-electron contributions are treated separately. Since the nucleus is treated as point-like on the atomic scale, one has
\begin{equation}
\sum_{IJ} P_{IJ}(r)
=
\frac{N_P}{2}\left(N_P-1\right)\delta(r),
\end{equation}
where $I,J$ label nuclei. For the nucleus-electron term, each nucleus contributes equally to the electronic radial distribution, hence
\begin{equation}
\sum_{iJ} P_{iJ}(r)
=
N_P \sum_i R_i(r)
=
N_P \sum_i 4\pi r^2 \rho_i(r),
\end{equation}
where $i$ labels electrons, $\rho_i(r)$ is the one-electron radial density, and $R_i(r)=4\pi r^2\rho_i(r)$. For the electron-electron term, one separates the diagonal and off-diagonal contributions:
\begin{equation}
\sum_{ij} P_{ij}(r)
=
\sum_{i\neq j} P_{ij}(r) + N_e \delta(r).
\end{equation}
The diagonal part gives the contact contribution $N_e\delta(r)$, while the off-diagonal part is described by the pair distribution
\begin{equation}
P(r)\equiv \sum_{i\neq j} P_{ij}(r),
\end{equation}
normalized as
\begin{equation}
\int_0^\infty dr\, P(r)
=
\frac{N_e}{2}(N_e-1).
\end{equation}
Using these relations in the general rate formula
\begin{equation}
\frac{d\Gamma}{dE} = \frac{1}{E}\bigg[
\frac{N_{P}}{2}\left( N_{P} -1\right) -N_PN_e + N_e + \frac{N_{e}}{2}( N_{e} -1)
\bigg]
\end{equation}
for a neutral atom, $N_P=N_e$, and the rate vanishes, as expected.
\section{Sensitivity to relativistic effects}
\label{sec:relativistic}
\begin{figure}[h!]
    \centering
    \includegraphics[width=\columnwidth]{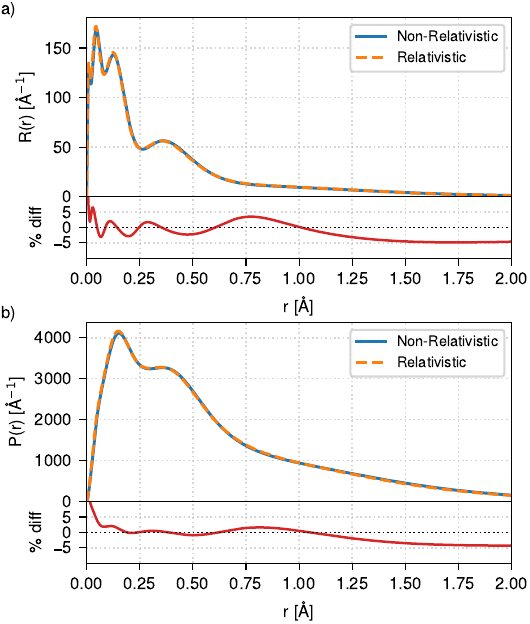}
    \caption{Sensitivity of the RDFs to relativistic effects in Xe for the nuclear-electron (a) and the electron-electron (b) RDFs. Each plot compares the RDF obtained using nonrelativistic and relativistic wavefunctions, with the lower panels report the relative percentage difference between the two calculations.}
    \label{fig:xe_sensitivity}
\end{figure}

To assess the impact of relativistic effects on the present framework, we compare the RDFs obtained using nonrelativistic and relativistic wavefunctions for Xe, the heaviest atom considered in this work. As shown in Figs.~\hyperref[fig:xe_sensitivity]{\ref*{fig:xe_sensitivity}(a)-\ref*{fig:xe_sensitivity}(b)}, the two calculations are in good agreement over the entire range of interparticle distances. Relative differences remain below approximately $5\%$ and are confined to specific regions of the RDFs, while the overall distributions are essentially unchanged. These results indicate that relativistic effects provide only a minor correction within the present framework, justifying the use of nonrelativistic wavefunctions throughout this work.
\vskip20pt
%
%
\FloatBarrier
\bibliographystyle{apsrev4-2}
\bibliography{collapse_rdf}
\end{document}